# An Optimal Method to Combine Results from Different Experiments

By Theodore P. Hill and Jack Miller

Georgia Institute of Technology and Lawrence Berkeley National Laboratory

hill@math.gatech.edu, miller@lbl.gov

When different experiments measure the same unknown quantity, such as Planck's constant, how can their results be consolidated in an unbiased and optimal way? Is there a good method to combine data from experiments that may differ in time, geographical location, and methodology and even in underlying theory?

The consolidation of data from different sources can be particularly vexing in the determination of the values of the fundamental physical constants. For example, the U.S. National Institute of Standards and Technology (NIST) recently reported "two major inconsistencies" in some measured values of the molar volume of silicon $V_m(Si)$ and the silicon lattice spacing $d_{220}$, leading to a recommended *ad hoc* factor of 1.5 increase in the uncertainty in the value of Planck's constant $h$ [MTN1, p. 54, MTN2]. (One of those two inconsistencies has subsequently been resolved [M].)

But input data distributions that happen to have different means and standard deviations are not necessarily "inconsistent" or "incoherent" [D, p 2249]. If the various input data are all normally or exponentially distributed, for example, then every interval centered at the unknown positive true value has a positive probability of occurring in every one of the independent experiments. Ideally, of course, *all* experimental data, past as well as present, should be incorporated into the scientific record. But in the case of the fundamental physical constants, for instance, this could entail listing scores of past and present experimental datasets, each of which includes results from hundreds of experiments with thousands of data points, for *each one of the fundamental constants*. Most experimentalists and theoreticians who use Planck's constant, however, need a concise summary of its current value rather than the complete record. Having the mean and estimated standard deviation (e.g. via weighted least squares) does give some information, but without any knowledge of the distribution, knowing the mean within two standard deviations is only valid at the 75% level of significance, and knowing the mean within four standard deviations is not even significant at the standard 95% confidence level. Is there an objective, natural and optimal method for consolidating several input-data distributions into a single posterior distribution $P$, without ad hoc adjustments or arbitrary assignment of weights to the input data sets? This article proposes such a method.

First, it is useful to review some of the shortcomings of standard methods for consolidating data from several different input distributions. For simplicity, consider the case of only two different experiments in which independent laboratories Lab I and Lab II measure the value of the same quantity. Lab I reports its results as a probability distribution $P_1$ (e.g. via an empirical histogram or probability density function), and Lab II reports its findings as $P_2$.

## Averaging the Probabilities

One common method of consolidating two probability distributions is to simply average them - for every set of values $A$, set $P(A) = (P_1(A) + P_2(A))/2$. If the distributions both have densities, for example, averaging the probabilities results in a probability distribution with density the average of the two input densities (Figure 1). This method has several significant disadvantages. First, the mean of the resulting distribution $P$ is always exactly the average of the means of $P_1$ and $P_2$, independent of the relative accuracies or variances of each. (Recall that the variance is the square of the standard deviation.) But if Lab I performed twice as many of the same type of trials as Lab II, the variance of $P_1$ would be half that of $P_2$, and it would be unreasonable to weight the two respective empirical means equally.

A second disadvantage of the method of averaging probabilities is that the variance of $P$ is always *at least as large* as the minimum of the variances of $P_1$ and $P_2$ (see Figure 1), since $\mathrm{var}(P) = (\mathrm{var}(P_1) + \mathrm{var}(P_2))/2 + (mean(P_1) - mean(P_2))^2/4$. If $P_1$ and $P_2$ are nearly identical, however, then their average is nearly identical to both inputs, whereas the standard deviation of a reasonable consolidation $P$ should be strictly less than that of both $P_1$ and $P_2$. The method of averaging probabilities completely ignores the fact that two laboratories independently found nearly the same results. Figure 1 also shows another shortcoming of this method - with normally-distributed input data, it generally produces a multimodal distribution, whereas one might desire the consolidated output distribution to be of the same general form as that of the input data - normal, or at least unimodal. Generalizing this method to allow biased (unequal) weights has the same drawbacks, and the additional problem of assigning and justifying the unequal weights.

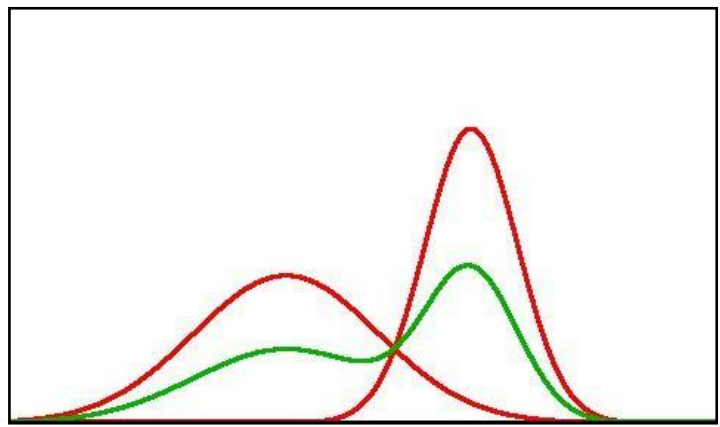

**Figure 1. Averaging the Probabilities.**

**(Green curve is the average of the red (input) curves. Note that the variance of the average is larger than the variance of either input.)**

**Averaging the Data**

Another common method of consolidating data - one that does preserve normality - is to average the underlying input data itself. That is, if the result of the experiment from Lab I is a random variable $X_1$ (i.e. has distribution $P_1$) and the result of Lab II is $X_2$ (independent of $X_1$, with distribution $P_2$), take $P$ to be the distribution of $(X_1 + X_2)/2$. As with averaging the distributions, averaging the data also results in a distribution that always has exactly the average of the means of the two input distributions, regardless of the relative accuracies of the two input data-set distributions (see Figure 2). With this method, on the other hand, the variance of $P$ is *never larger* than the maximum variance of $P_1$ and $P_2$ (since $\mathrm{var}(P) = (\mathrm{var}(P_1) + \mathrm{var}(P_2))/4$), whereas some input data distributions that differ significantly should reflect a higher uncertainty. A more fundamental problem with this method is that in general it requires averaging data obtained using very different and even indirect methods, including for example the watt balance and x-ray and optical interferometer measurements used in part to obtain the 2006 CODATA recommended value for Planck's constant.

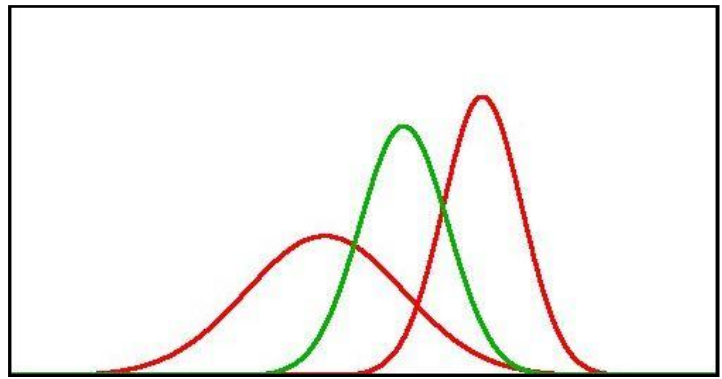

**Figure 2. Averaging the Data**

**(Green curve is the average of the red data. Note that the mean of the averaged data is exactly the average of the means of the two input distributions, even though they have different variances)**

## New Method – Conflating the Data

An alternative method for consolidating different input distributional data, called the *conflation* of distributions (designated with the symbol "&" to suggest consolidation of $P_1$ *and* $P_2$) has none of these disadvantages. Conflation is easy to calculate and visualize, and has many useful properties. If the input distributions $P_1, P_2, ...P_n$ all have densities, then the conflation $\&(P_1, P_2, ..., P_n)$ of $P_1, P_2, ...P_n$ is the probability distribution with density that is the normalized product of the input densities. That is, if the densities of $P_1, P_2, ...P_n$ are $f_1, f_2, ..., f_n$, respectively, then $\&(P_1, P_2, ..., P_n)$ is the probability distribution with density (see Figure 3)

$$f(x) = \frac{f_1(x) \times f_2(x) \times ... \times f_n(x)}{\int_{-\infty}^{\infty} f_1(t) \times f_2(t) \times ... \times f_n(t) dt} \;.$$

(If the denominator is zero or is infinite, the definition is slightly different, and for discrete input distributions, the analogous definition is the normalized product of the probability mass functions – see [H] for details.) The conflation of distributions has a natural heuristic and practical interpretation – gather data from the independent laboratories sequentially and simultaneously, and record the values only when the laboratories (nearly) agree.

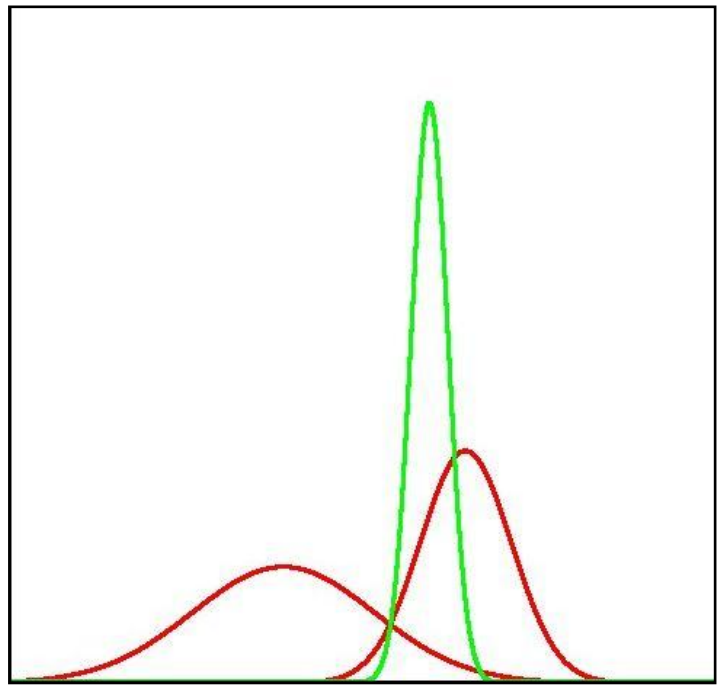

**Figure 3. Conflating Distributions**

**(Green curve is the conflation of red curves. Note that the mean of the conflation is closer to the mean of the input distribution with smaller variance, i.e. with greater accuracy)**

It may at first glance seem counterintuitive that the conflation of two relatively broad distributions can be a much narrower one (Figure 3). However if both measurements are assumed equally valid, then the true value should lie in the overlap region between the two distributions with high probability. Looking at it statistically, if one lab makes 50 measurements and another lab makes 100, then the standard deviations of their resulting distributions will usually be different. If the labs' methods are different, with different systematic errors, or their methods rely on different fundamental constants with different uncertainties, then the means will likely be different too. But the bottom line is that the total of 150 valid measurements is substantially greater than either lab's data set, so the standard deviation should indeed be smaller.

**Advantages of Conflation**

Conflation has significant practical and mathematical advantages. It is easy to calculate and visualize, and easy to update (simply conflate the latest input with the overall conflation of past inputs). As was shown in Figure 3, the mean of the conflation gives more weight to means of input distributions arising from more accurate experiments (recall that the methods of averaging the probabilities and averaging the data both result in a distribution whose mean is exactly the average of the input means; Figures 1 and 2.)

Conflations of normal (Gaussian) distributions are always normal (see Figure 3, and red curve in Figure 4B), and coincide with the classical weighted least squares method, hence yielding BLUE (Best Linear Unbiased Estimators) and MLE (Maximum Likelihood) estimators (cf. [A],[RS]). Many of the other important classical families of distributions, including gamma, beta, uniform, exponential, Pareto, Laplace, Bernoulli, zeta and geometric families, are also preserved under conflation.

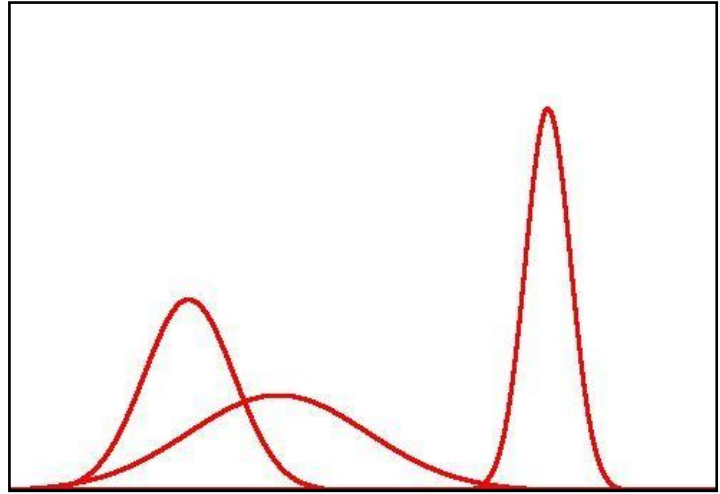

**Figure 4A. Three Input Distributions**

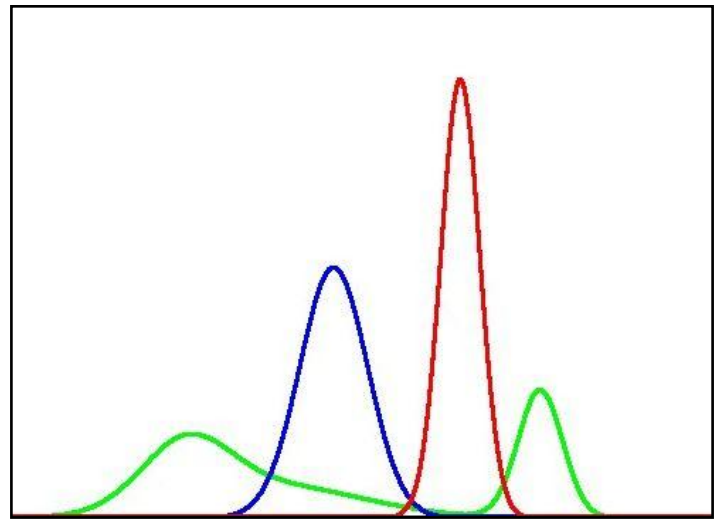

**Figure 4B. Comparison of Averaging Probabilities, Averaging Data, and Conflating**

**(Green curve is average of the three input distributions in Figure 4A, blue curve is average of the three input datasets, and red curve is the conflation.)**

Most importantly, however, the conflation of several distributions is optimal in three significant mathematical respects. First, the conflation is the unique distribution that minimizes the maximum loss of the *Shannon Information*, a classical formal measure of information. When the goal is to consolidate information from several (input) distributions into a single (output) distribution, replacing several distributions by a single distribution will clearly result in some loss of information, however that is defined. The Shannon Information (also called the *surprisal*, or *self-information*), $S_P(A)$, obtained by observing an event *A* in an experiment *P* is $S_P(A) = -\log_2 P(A)$ (so the smaller the value of $P(A)$, the greater the information or surprise). The (combined) Shannon Information *of $P_1, P_2, ..., P_n$* for the event *A*, is

$$S_{\{P_1,\ldots,P_n\}}(A) = \sum_{i=1}^{n} S_{P_i}(A) = -\log_2 \prod_{i=1}^{n} P_i(A)$$, and the loss between the Shannon Information of a distribution *P and* $P_1,\ldots,P_n$ for the event *A* is $S_{\{P_1,\ldots,P_n\}}(A) - S_P(A)$. Thus the loss of information depends on the outcome *A* observed. The conflation of $P_1, P_2, \ldots, P_n$ is the unique probability distribution *P* that makes the maximum loss of Shannon Information, among all possible observed outcomes *A*, as small as possible.

The conflation of the distributions is also the unique probability distribution that makes the variation of these likelihood ratios as small as possible. In classical hypotheses testing, a standard technique to decide from which of *n* known distributions given data actually came is to maximize the likelihood ratios, that is, the ratios of the probability density or probability mass functions. Analogously, when the objective is how best to consolidate data from those input distributions into a single (output) distribution *P*, one natural criterion is to choose *P* so as to make the ratios of the likelihood of observing *x* under *P* to the likelihood of observing *x* under *all* of the (independent) distributions $\{P_i\}$ as close as possible.

The conflation of the distributions is also the unique probability distribution that preserves the proportionality of likelihoods. A criterion similar to likelihood ratios is to require that the output distribution *P* reflect the relative likelihoods of identical individual outcomes under the $\{P_i\}$. For example, if the likelihood of all the experiments $\{P_i\}$ observing the identical outcome *x* is twice that of the likelihood of all the experiments $\{P_i\}$ observing *y*, then *P(x)* should also be twice as large as *P(y)*.

Conflation has one more advantage over the methods of averaging probabilities or data. In practice, assumptions are often made about the form of the input distributions, such as an assumption that underlying data is normally distributed [MTN1]. But the true and estimated values for Planck's constant are clearly never negative, so the underlying distribution is certainly not truly normally distributed – more likely it is truncated normal. Using conflations, the problem of truncation essentially disappears – it is automatically taken into account. If one of the input distributions is summarized as a true normal distribution, and the other excludes negative values, for example, then the conflation will exclude negative values, as is seen in Figure 5.

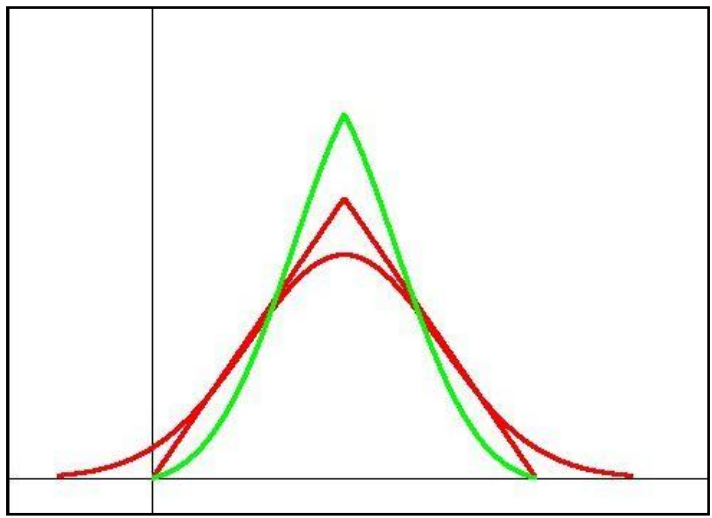

**Figure 5.** **(Green curve is the conflation of red curves. Note that the conflation has no negative values, since the triangular input had none.)**

**Example with {220} Lattice Spacing Measurements**

The input data used to obtain the CODATA 2006 recommended values and uncertainties of the fundamental physical constants includes the measurements and inferred values of the absolute {220} lattice spacing of various silicon crystals used in the determination of Planck's constant and the Avogadro constant. The four measurements came from three different laboratories, and had values 192,015.565(13), 192,015.5973(84), 192,015.5732(53) and 192,015.5685(67), respectively [MTN2, TableXXIV], where the parenthetical entry is the uncertainty. The CODATA Task Force viewed the second value as "inconsistent" with the other three (see red curves in Figure 6) and made a consensus adjustment of the uncertainties. Since those values "are the means of tens of individual values, with each value being the average of about ten data points" [MTN2], the central limit theorem suggests that the underlying datasets are approximately normally distributed as is shown in Figure 6 (red curves). The conflation of those four input distributions, however, requires no consensus adjustment, and yields a value essentially the same as the final CODATA value, namely, 192,015.5762 [MTN2, Table LIII], but with a much smaller uncertainty. Since uncertainties play an important role in determining the *values* of the related constants via weighted least squares, this smaller, and theoretically justifiable, uncertainty is a potential improvement to the current accepted values.

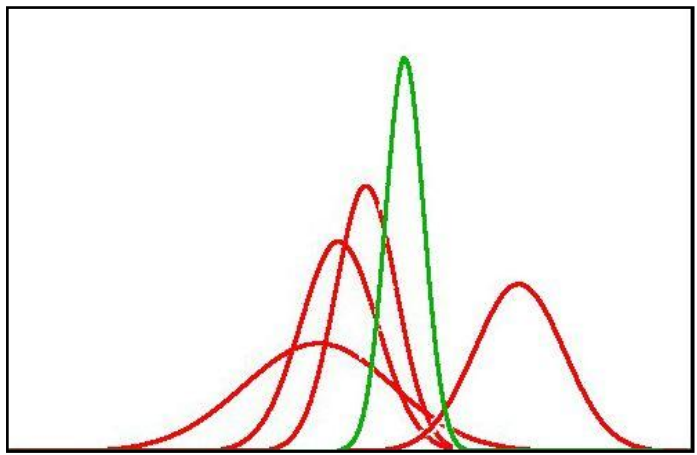

**Figure 6. (The four red curves are the distributions of the four measurements of the {220} lattice spacing underlying the CODATA 2006 values; the green curve is the conflation of those four distributions, and requires no ad hoc adjustment.)**

## Conclusion

The conflation of several input-data distributions is a probability distribution that summarizes the data in an optimal and unbiased way. The input data may already be summarized, perhaps as a normal distribution with given mean and variance, or may be the raw data themselves in the form of an empirical histogram or density. The conflation of these input distributions is easy to calculate and visualize, and affords easy computation of sharp confidence intervals. Conflation is easy to update, is the unique minimizer of loss of Shannon information , the unique minimal likelihood ratio consolidation and the unique proportional consolidation of the input distributions. Conflation of normal distributions is always normal, and conflation preserves truncation of data. Perhaps the method of *conflating* input data will provide a practical and simple, yet optimal and rigorous method to address the basic problem of consolidation of data.

## Acknowledgement

The authors are grateful to Dr. Peter Mohr for enlightening discussions regarding the 2006 CODATA evaluation process.